\documentclass[12pt, oneside]{article}

\usepackage[utf8]{inputenc}
\usepackage[T1]{fontenc}
\usepackage{mathptmx}
\usepackage{setspace} 
\usepackage[margin=1in]{geometry}
\usepackage{apacite}

\usepackage[document]{ragged2e}
\usepackage{graphicx}
\usepackage{float}
\usepackage{amsthm}
\usepackage{subcaption}
\usepackage{booktabs}
\usepackage{amssymb}
\usepackage{bm}
\usepackage{bbm}
\usepackage{enumitem}
\usepackage{amsmath}
\theoremstyle{plain}

\usepackage[labelfont=bf]{caption}

\usepackage{natbib}
 \bibpunct[, ]{(}{)}{,}{a}{}{,}%

\DeclareMathOperator*{\argmin}{arg\,min}

\usepackage{balance}
\usepackage[pdflang={en-US},pdftex]{hyperref}
\usepackage{color}
\usepackage{textcomp}
\usepackage{tabularx}
\usepackage{parskip}

\usepackage{algpseudocode}
\usepackage[section]{placeins}

\graphicspath{{figures/}}

\usepackage[small,compact]{titlesec}
\titleformat*{\section}{\centering\bfseries\MakeUppercase}
\titleformat*{\subsection}{\centering\bfseries}

\let\oldbibliography\bibliography
\renewcommand{\bibliography}[1]{%
  {\fontfamily{ptm}\selectfont
   \oldbibliography{#1}}
}

\usepackage{siunitx}
\usepackage{wrapfig}
\usepackage{amsfonts}
\usepackage{makecell}
\usepackage{array}

\title{PRISM: A Design Framework for Open-Source Foundation Model Safety}
\date{}

\author{Terrence Neumann\\
        McCombs School of Business\\
        University of Texas at Austin\\
        \and
        Bryan Jones\\
        School of Law\\
        University of Texas at Austin}

\begin{document}

\maketitle
\singlespacing
\begin{abstract}
The rapid advancement of open-source foundation models has brought transparency and accessibility to this groundbreaking technology. However, this openness has also enabled the development of highly-capable, unsafe models, as exemplified by recent instances such as WormGPT and FraudGPT, which are specifically designed to facilitate criminal activity. As the capabilities of open foundation models continue to grow, potentially outpacing those of closed-source models, the risk of misuse by bad actors poses an increasingly serious threat to society. This paper addresses the critical question of how open foundation model developers should approach model safety in light of these challenges. Our analysis reveals that open-source foundation model companies often provide less restrictive acceptable use policies (AUPs) compared to their closed-source counterparts, likely due to the inherent difficulties in enforcing such policies once the models are released. To tackle this issue, we introduce PRISM, a design framework for open-source foundation model safety that emphasizes \textbf{P}rivate, \textbf{R}obust, \textbf{I}ndependent \textbf{S}afety measures, at \textbf{M}inimal marginal cost of compute. The PRISM framework proposes the use of modular functions that moderate prompts and outputs independently of the core language model, offering a more adaptable and resilient approach to safety compared to the brittle reinforcement learning methods currently used for value alignment. By focusing on identifying AUP violations and engaging the developer community in establishing consensus around safety design decisions, PRISM aims to create a safer open-source ecosystem that maximizes the potential of these powerful technologies while minimizing the risks to individuals and society as a whole. \end{abstract}

\clearpage

\section{Introduction}
The rapid advancement of open-source foundation models has led to the widespread release of powerful tools that can be easily fine-tuned for various purposes, both beneficial and malicious. While open-source development is celebrated for its accessibility, transparency, and enhanced privacy for end-users compared to API-restricted access, it also presents significant challenges in terms of model safety. Recent examples like WormGPT and FraudGPT, which facilitate cybercrime by fine-tuning open large language models, underscore the urgent need for robust safety measures in open-source AI development.

One of the primary issues is the difficulty open-source developers face in monitoring and enforcing their acceptable use policies (AUPs) \citep{klyman2024aups}. This lack of oversight makes the models vulnerable to misuse by malicious actors, posing a considerable risk to society. Our research, which includes an analysis of a dataset compiling AUPs from various foundation model companies, reveals that open-source developers tend to provide fewer policies across many categories of potential harms compared to their closed-model counterparts. This discrepancy may stem from the inherent challenges in enforcing such policies once the models are released or from the open-source philosophy that values accessibility and innovation over strict limitations.

These enforcement challenges and the resulting policy gaps raise a critical question: \emph{how should open foundation model developers approach model safety}? To address this, we introduce PRISM, a forward-looking framework designed to guide open-source foundation model development. PRISM stands for \textbf{P}rivate, \textbf{R}obust, \textbf{I}ndependent \textbf{S}afety at \textbf{M}inimal marginal cost of compute. This framework emphasizes the importance of integrating robust safety measures that do not compromise end-user privacy and are not dependent on the specific language model architecture. By focusing on these principles, PRISM aims to create a safer open-source environment without imposing significant additional computational costs on developers or users.

In this paper, we begin by providing a background section on open and closed foundation model development, covering key definitions, notions of model safety and vulnerabilities, explaining the capability gaps between open and closed foundation models, and discussing how AUPs are implemented and enforced to self-regulate the use of foundation models. In the next section, we propose the development of a large language model that embodies the PRISM principles. By adopting the PRISM design principles, we argue that it is possible to achieve utility improvements for both end-users and society-at-large. Implementing privacy-preserving techniques, enhancing model robustness, and ensuring that safety measures are model-independent and cost-effective can help mitigate the risks associated with open foundation models. Our goal is to foster a responsible open-source community that balances the benefits of accessibility and transparency with the imperative of protecting against misuse and harm.

\section{Background \& Motivation}
\subsection{A Working Definition of Open and Closed Foundation Models}

A foundation model is ``any model that is trained on broad data (generally using self-supervision at scale) that can be adapted (e.g., fine-tuned) to a wide range of downstream tasks'' \citep[p.3]{bommasani2021opportunities}. Open foundation model development is an evolving concept, as many publications have described a ``gradient'' of openness \citep{solaiman2023gradient}, with various levels of transparency into model design. For instance, foundation model developers can open-source the model weights, but not the model code or training data, leaving questions as to the ultimate transparency or ``openness'' of the foundation model.

We define an \textbf{open foundation model} as \emph{a foundation model in which the pre-trained weights are available for end-users to download}, acknowledging that some may define openness differently. Conversely, we define a \textbf{closed foundation model} as \emph{a foundation model in which direct end-user access to model weights is restricted}, with model access achieved via third-party API requests.

\subsection{Open and Closed Foundation Model Safety and Vulnerabilities}

Foundation models, regardless of whether they are open or closed, present challenges to society. Propaganda generated by large language models (LLMs) can be more convincing than human-created content \citep{AIdisinfoSpitale2023}, and adversarial actors can save costs using LLMs for influence operations \citep{musser2023cost}. There is also a risk of proliferating child sexual abuse material via image generation models \citep{thiel2023generative}.

There has been significant debate about the safety of open foundation models in particular, as their advanced capabilities and lack of centralized governance present substantial risks of misuse. Some argue for more restricted release, citing threats like influence operations, surveillance, scamming, cyber-attacks, and the development of banned weapons \citep{seger2023opensourcing}. Others argue that the threats are overblown and that open models offer societal benefits such as accelerating science, allowing broader definitions of acceptable behavior, and enhancing economic innovation \citep{kapoor2024societal}. 

To mitigate these threats, developers have attempted to align models with human values through reinforcement learning \citep{bai2022training}. However, the learned reward functions often fail to capture the subtlety of human values \citep{bommasani2021opportunities}, especially when values conflict. This alignment approach is vulnerable to ``jailbreak'' attacks like prompt injection \citep{chao2023jailbreaking} and malicious fine-tuning \citep{qi2023finetuning}, which can drastically increase unsafe outputs. The brittleness of reinforcement learning for safety is evident, as only 3\% of neurons are uniquely responsible for safe outputs in some models \citep{wei2024assessing}. Real-world examples like WormGPT and FraudGPT\footnote{\url{https://thehackernews.com/2023/07/new-ai-tool-fraudgpt-emerges-tailored.html}}\footnote{\url{https://thehackernews.com/2023/07/wormgpt-new-ai-tool-allows.html}} demonstrate this vulnerability. Such models have been fine-tuned to provide criminal or otherwise harmful outputs, ranging from hacking instructions, malicious codes, to detailed plans on how to conduct advanced phishing schemes. Some have described ``catastrophic robustness failures'' of model alignment and safety as among the greatest challenges facing open foundation model developers \citep{bommasani2021opportunities}.

This paper presents an alternative safety design that identifies unsafe prompts or outputs via independent models, rather than relying on the complex process of reinforcement learning to align with diverse human values.

\subsection{The (Narrowing) Capability Gap Between Open and Closed Foundation Models}

\begin{figure*}
 \centering
    \includegraphics[width=\textwidth]{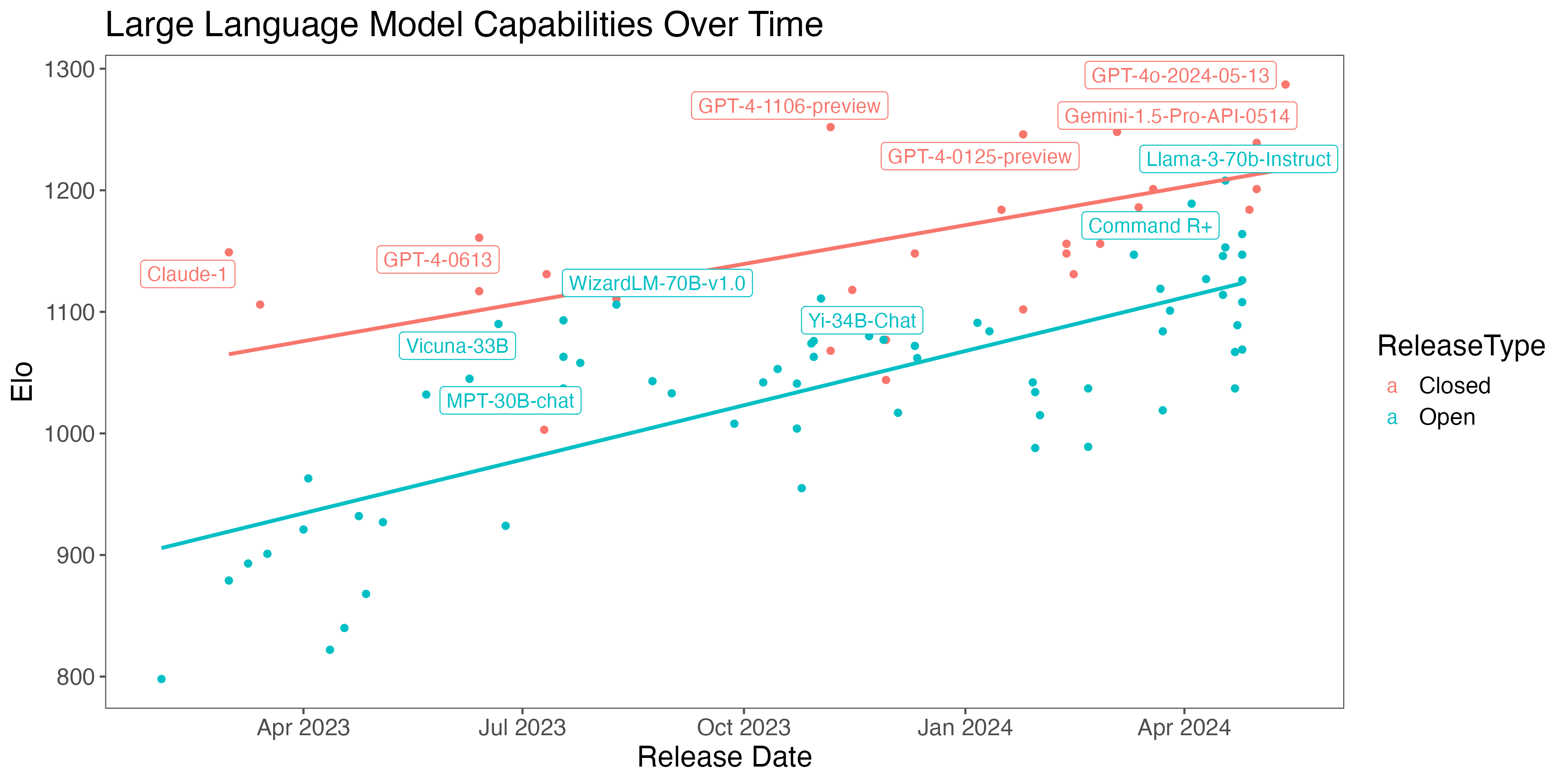}
    \caption{The evolving capabilities of open- and closed-source language models over time. Open-source models are improving at a rate at least as fast as closed-source models. Source: \citep{chiang2024chatbot}}
    \label{fig:closed_open_capability}
\end{figure*}

The capabilities of open foundation models are advancing at a rate that may outpace that of closed models, driven by the collective efforts of a global community of researchers and developers. \citet{chiang2024chatbot} have gathered millions of rows of human feedback data in which model output from many models are compared pairwise (e.g., ``Do you prefer the output of Model A or Model B?'') and models are ranked via a method that tabulates wins and losses similar to how chess player rankings are calculated \citep{boubdir2023elo}. By compiling these scores for nearly 100 model releases over the past year and identifying the models as either open- or closed-source, we can determine the rate of improvement of both model types over time. From a simple regression analysis of this data, we see that the coefficient on the daily rate of capability improvement of open-source models is greater ($\beta_{open} = 0.485$), but not statistically different, than the daily rate of capability improvement of closed-source models ($\beta_{closed} = 0.347$). See Figure \ref{fig:closed_open_capability} and Table \ref{tab:regression}.

\begin{table}[h]
\centering
\begin{tabular}{l|c}
\hline
& Estimate (Std. Error) \\
\hline
\hline
(Intercept) & $1055\textsuperscript{***}$ \\
& $(28.2)$ \\
Release Date & $0.3470\textsuperscript{***}$ \\
& $(0.080)$ \\
Open-Source & $-149.9\textsuperscript{***}$ \\
& $(32.5)$ \\
Release Date:Open-Source & $0.1383$ \\
& $(0.095)$ \\
\hline
\multicolumn{2}{l}{$R^2 = 0.698$} \\
\multicolumn{2}{l}{\footnotesize{$^{***}p<0.001$, $^{**}p<0.01$, $^{*}p<0.05$}} \\
\end{tabular}
\caption{Results for regression of model Elo score on Release Date (indicating the daily rate of capability improvement), whether they are open-source, and the interaction between the two.}
\label{tab:regression}
\end{table}

If this accelerated rate of improvement persists, open-source models may well become the predominant mode of foundation model development and usage, particularly among businesses with the resources to host and customize these models. The cost-effectiveness of open-source models, which eliminates the need for expensive per-inference fees, is a significant advantage for businesses seeking to leverage advanced AI technologies. 

The AI landscape is adversarial and dynamic, characterized by attackers intent on causing societal harm and the constantly evolving capabilities of foundation models, their defenses, and the attackers. Thus, as the capabilities of open-source foundation models increase, so too does the potential for misuse. The advancement in capabilities not only amplifies existing threat vectors but also introduces new ones that may have been previously unconsidered. This dynamic nature of emerging threats necessitates a continuous and adaptable approach to safety and control measures. The open-source nature of these models, while promoting innovation and accessibility, simultaneously poses a challenge in maintaining robust oversight and ensuring alignment with societal values. As such, the process of safeguarding these models must be iterative, involving regular updates and enhancements to security protocols, threat assessments, and ethical guidelines. 

\subsection{Self-Governance via Acceptable Use Policies of Foundation Models}

While some government guidelines exist limiting or restricting the use of foundation models to unsafe applications, in general, most end-user restrictions are not comprehensively defined by governments at this point. For instance, the FTC has issued warnings against the use of AI to make false or unsubstantiated claims, emphasizing that companies must ensure their AI-generated content is truthful and not misleading \citep{atleson2023ai}. However, broader regulatory frameworks specifically addressing the deployment and use of foundation models in various sectors, including the public sector \citep{adalovelace2023foundation}, are still evolving. 

In response to potential misuse of their models, foundation model companies often make end-users agree to specific acceptable use policies (AUPs) that define restrictions on model usage. These AUPs cover often cover a wide range of potential societal harms, ranging from the production of mis- and dis-information, to privacy violations, to encouraging self-harm or other dangerous behavior. However, enforcement of acceptable use policies is difficult in practice. As recently as last year, researchers demonstrated that there was little difference between closed- and open-source large language models in their defenses against malicious fine-tuning attacks \citep{qi2023finetuning}. However, closed-model companies theoretically have greater centralized means to dynamically update ``content moderation'' style filters as new threats emerge, while open foundation model developers currently forgo monitoring and updating the system once it is in the hands of end-users.

\begin{figure*}
 \centering
    \includegraphics[width=\textwidth]{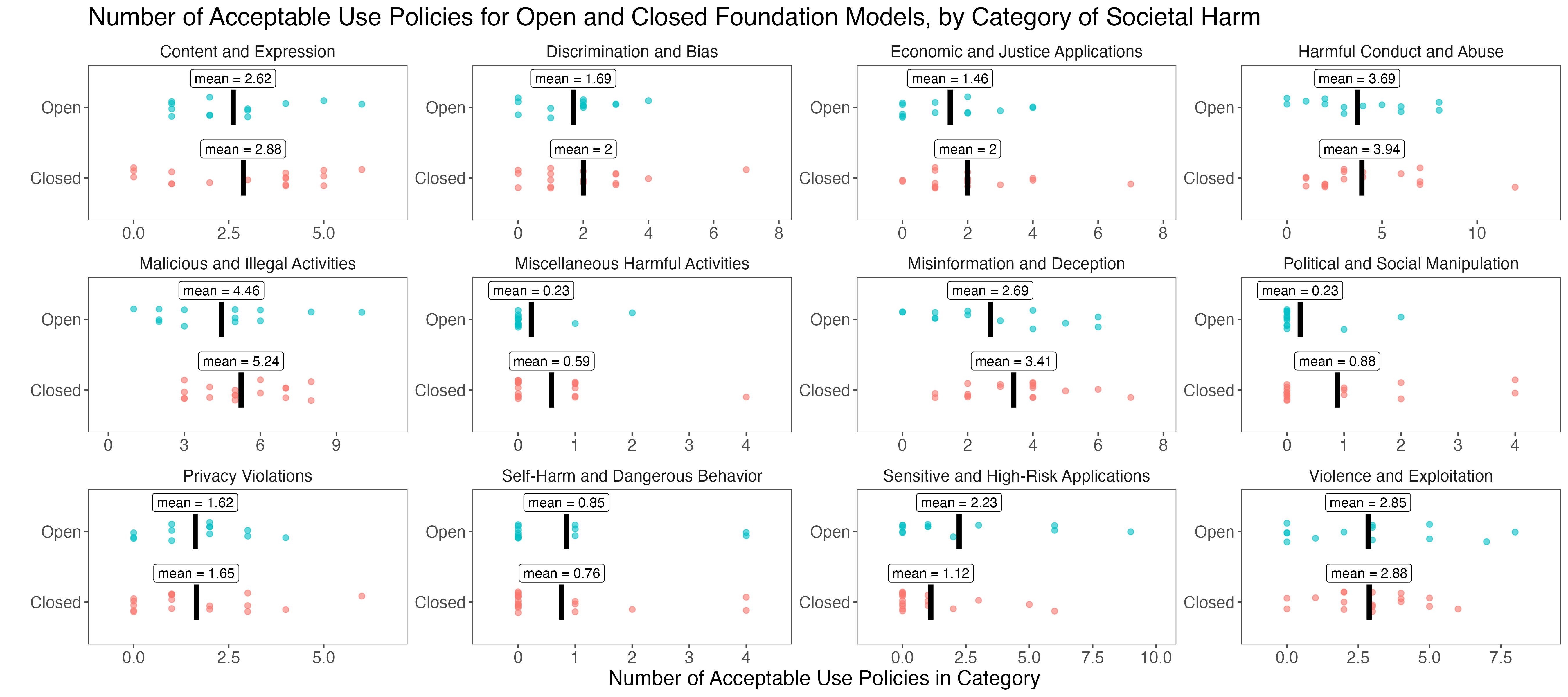}
    \caption{This chart shows the mean number of acceptable use policies (AUPs) for open- and closed-source foundation models across a variety of categories of potential harm. In many cases, closed-source models seem to provide more terms of use intended to mitigate societal harm.}
    \label{fig:closed_open_tou}
\end{figure*}

In order to analyze whether open- or closed-source models implement more restrictive AUPs, we examine a new AUP dataset \citep{klyman2024aups}, adding a variable to distinguish companies as either open-source or closed-source developers based on publicly available information. We then compared the number of individual policies addressing different types of harm to society using the taxomy provided by \citet{klyman2024aups} as inspiration. Our analysis reveals that, in most harm categories of AUPs, closed-source models have a higher average number of policies, indicating more restrictive usage. For instance, we see higher average values for the number of AUPs related to: challenging types of content and expression ($\mu_{open} = 2.61$; $\mu_{closed} = 2.88$); discrimination and bias ($\mu_{open} = 1.69$; $\mu_{closed} = 2.00$); sensitive economic and justice applications ($\mu_{open} = 1.46$; $\mu_{closed} = 2.00$); misinformation and deception ($\mu_{open} = 2.69$; $\mu_{closed} = 3.41$); and malicious and illegal activities ($\mu_{open}=4.46$; $\mu_{closed} = 5.23$). However, we see open source foundation models having higher number of AUPs related to: encouraging self-harm or other dangerous behavior ($\mu_{open} = 0.85$; $\mu_{closed}=0.76$); and sensitive and high-risk applications, which include numerous national security threats ($\mu_{open}=2.23$; $\mu_{closed} = 1.11$). See Figure \ref{fig:closed_open_tou} for a visualization of this data.

The differences in AUPs between open-source and closed-source models may stem from the inherent principles of openness and accessibility in open-source development, which prioritize innovation and collaboration over stringent usage limitations. Additionally, open-source developers recognize the practical challenges of enforcing terms of use once models are released, given their free accessibility and potential for modification. As a result, they may opt for less comprehensive AUPs and rely on decentralized, voluntary self-enforcement by the user community. In contrast, closed-source model developers have more centralized control and resources to implement and enforce rigorous usage restrictions. The disparity in AUP comprehensiveness reflects these differing priorities and practicalities in model governance and safety management between open-source and closed-source models.

\section{The PRISM Design Framework for Open-Source Foundation Model Safety}
The previous section highlighted critical elements in the open-source foundation model safety debate. We noted that reinforcement learning, the primary paradigm for safety alignment, is vulnerable to attacks optimized when attackers access model weights. We also discussed the rapidly expanding capabilities of open-source models, potentially outpacing closed-source counterparts, which amplifies both their potential uses and misuses. Furthermore, we showed that open-source models generally have fewer AUPs across various societal harm categories, either due to their "openness" ethos or the acknowledged challenges in monitoring model usage.

In response to these concerns, we propose an innovative open-source LLM that tackles the practical challenges of open foundation model safety head-on. Our groundbreaking model embodies three core principles: (1) privacy, (2) robust, model-independent safety, and (3) minimizing the marginal cost of compute. In the following sections, we will delve into how each component delivers utility gains for end-users and society as a whole, paving the way for a safer and more responsible open-source AI landscape.

\subsection{Large Language Model Formulation}

In this section, we formulate a safety mechanism for a language model $f$ that embodies the PRISM principles. Let a pre-trained large language model be denoted as a function $f$ which, during inference, takes an input sequence of tokens (a prompt) $x = (x_1, x_2, \dots, x_N)$ and has outputted tokens $y_1, y_2, ..., y_j$ so far. The function $f$ outputs a probability distribution over the vocabulary $V$ for the next output token: $f(x, y_1, y_2, \dots, y_j) = P(y_{j+1} = v | x, y_1, y_2, \dots, y_j)$ for $v \in V$. By iteratively selecting the next token with the highest probability, a completed output sequence $Y = (y_1, y_2, \dots, y_J)$ is produced.

Rather than directly modifying the probability distribution responsible for producing the output sequence $Y$, which is the outcome of current reinforcement learning approaches to safety alignment of foundation models, we instead propose modular ``interceptor'' functions that moderate prompts and outputs: $p(x)$ and $q(Y)$ respectively. Interceptor functions are language models trained to enforce the AUPs of a given model. These functions take the input sequence $x$ or the output sequence $Y$ as input and return a probability of the next token in the prompt or output being ``unsafe'': $p(x) = P(x_{i+1} \text{ is ``unsafe''} | x_1, x_2, \dots, x_i)$ and $q(Y) = P(y_{j+1} \text{ is ``unsafe''} | y_1, y_2, \dots, y_j)$. Then, a ``blocking'' function returns a binary sequence indicating which tokens are to be removed, given a probability threshold for being ``unsafe'': $b_{input}(x, p, \tau_{1}) = (b_{1}, b_{2}, \dots, b_{N})$ where $b_{i} = 1$ if $p(x_{i}) > \tau_{1}$ and $b_{i} = 0$ otherwise for $i = 1, 2, \dots, N$, and similarly, $b_{output}(Y, q, \tau_{2}) = (b_{1}, b_{2}, \dots, b_{J})$ where $b_{j} = 1$ if $q(y_{j}) > \tau_{2}$ and $b_{j} = 0$ otherwise for $j = 1, 2, \dots, J$. Lower values of $\tau$ indicate stricter moderation of prompts and outputs, and the inclusion of criteria as ``safe'' or ``unsafe'' will impact the underlying probability distribution.

Finally, numerous interceptor models may be capable of enforcing AUPs to an acceptable standard, and let these sets be defined as $P$ and $Q$. Choosing an inefficient or overly complex model can significantly add to computational expense, making this a non-starter for many end-users. Thus the objective of the system, holding the language model $f$ constant, could be characterized as: 

\begin{equation}
\begin{aligned}
& \argmin_{p \in P, q \in Q} \quad C(p, q) \\
& \text{subject to:} \\
& \quad U(p, q, \tau_{1}, \tau_{2}) - U_0 > 0 \\
\end{aligned}
\end{equation}

Where $C(p, q)$ represents the computational cost function associated with additional models $p$ and $q$, $U(p, q, \tau_{1}, \tau_{2})$ is the utility of the entire system (however that is defined by the developer), holding $f$ fixed.  $U_{0}$ is some target utility, perhaps defined by prior benchmarks. Thus, the objective is to minimize the cost of compute associated with $p$ and $q$ from the broader set of tested models $P$ and $Q$ subject to the constraints that model utility does not fall below a desired threshold.

\subsection{\underline{P}rivate}

$p$ and $q$ are not inherently privacy-preserving. In theory, end-user data could be collected and used for further training the system. However, as previously mentioned, a significant appeal of open foundation models is that they can be stored on premises and used with sensitive data that contain legal stipulations regarding user privacy. Therefore, training a robust $p$ and $q$ while maintaining end-user privacy is a challenge worth considering for open-source foundation model developers. 

One approach may be to host user hackathons and bounty programs, offering up cash incentives for spotting model vulnerabilities or potential attacks\footnote{\url{https://bughunters.google.com/about/rules/google-friends/5238081279623168/abuse-vulnerability-reward-program-rules}}. Additionally, this serves as an opportunity to make a developer organization's choice of $p$ and $q$ transparent as well, since it will not be relying on private end-user data. Because of the modular nature of the model, these can be refined and updated quickly and with little cost. Community engagement can help build consensus around challenging design decisions related to model safety.

Further, it is possible to maintain privacy while also enforcing a policy requiring end-users to maintain the most up-to-date interceptor models $p$ and $q$. This could be done, for instance, through licensing. An API could determine whether model weights for $p$ and $q$ have been updated in the central model repository as new threats emerge. If they have been updated, end-users would be required to update their local repository to reflect these changes. Failure to do so would void their license, rendering model inference impossible. Because $p$ and $q$ are independent from $f$, pushing updates to these models should not result in system-wide issues, as fine-tuning only affects $f$.

\paragraph{Utility Gains for End-Users}

For end-users, keeping the core language model ($f$) on-premises and only updating the smaller interceptor models ($p$ and $q$) as needed ensures that sensitive data remains private and is not shared with model developers or third parties. This is especially important for users in industries with strict data privacy regulations, which may curtail their ability to interact with closed-source model APIs \citep{sher2023privacy}. Additionally, privacy-as-a-policy helps mitigate many risks associated with data collection and retention that individual users face \citep{shahriar2023privacy}.

\paragraph{Utility Gains for Society-at-Large}

For society-at-large, the proposed model formulation reduces the risk of data misuse or unintended biases being introduced into the language model by keeping user data private and not relying on it for further training. This helps maintain trust in the technology and its applications \citep{bansal2015role}. The transparent development of interceptor models through user hackathons, posted bounties, and community feedback encourages the establishment of widely accepted safety standards and best practices, ensuring that the language model is being used responsibly and in line with societal values.

\subsection{\underline{R}obust \& \underline{I}ndependent}

As we previously mentioned, reinforcement learning is a particularly brittle approach to safety \citep{wei2024assessing}. Therefore, catastrophic robustness failures are among the greatest challenges facing open foundation model development \citep{bommasani2021opportunities}. 

In order to improve safety robustness, we introduce modular $p$ and $q$ functions that moderate input and output, which should provide advantages against both of the most common attacks - prompt injection and malicious fine-tuning. For instance, if an attacker attempts a prompt injection attack, they may try to access dangerous content by sneaking past defenses used when interpreting the input prompt. However, in this context, even if a prompt fools $p$, if the output is dangerous, it should be moderated by $q$, removing the unsafe tokens. Relatedly, models can be fine-tuned, perhaps to be malicious, but the prompts and outputs will still have to pass the scrutiny of $p$ and $q$. Therefore, the safeguards that enforce the AUPs are protected from malicious fine-tuning attempts. 

Switching from reinforcement learning to a more modular approach to AI safety may benefit the model's overall capabilities. For instance, research has demonstrated that a language models capabilities scale more linearly with dataset size and number of parameters when they are optimized to predict the next token  \citep{kaplan2020scaling}. Thus, as the amount of training data and model complexity increases, the model's performance on language tasks improves in a more predictable and consistent manner. By concentrating on this core objective of language modeling, the modular approach can potentially achieve better performance with less computational resources compared to reinforcement learning-based methods.

\paragraph{Utility Gains for End-Users}

For end-users, the introduction of modular interceptor functions $p$ and $q$ that moderate input and output improves safety robustness against common attacks such as prompt injection and malicious fine-tuning. This can prove extremely advantageous for businesses as end-users, who seek to limit their liability for AI generated content. Researchers have demonstrated that reinforcement learning approaches to model safety are so fragile that, with as few as 10 non-malicious examples (i.e. standard business training data) \citep{qi2023finetuning}, safety guardrails can accidentally be significantly disrupted. With this modular approach, end-users could reap the scaling benefits of a pure large language model \citep{kaplan2020scaling}, and know that the model safety is robust.

\paragraph{Utility Gains for Society-at-Large}

From a societal perspective, by providing a more resilient framework for enforcing AUPs and mitigating the risks associated with common attacks, the proposed formulation helps address public concerns about the potential misuse of open-source foundation models, including for fraud and crime \citep{seger2023opensourcing}. The modular nature of the safety mechanisms also allows for the incorporation of diverse perspectives and expertise in their development, such as the use of custom dictionaries and subject-matter expert hand-crafted datasets, further enhancing the robustness and adaptability of the system to evolving safety challenges.

\subsection{\underline{M}inimal Cost of Compute} Given the choice between two models with the same capabilities, the end-user is more likely to prefer a model with the smallest compute requirements. Less computationally expensive models will be cheaper and faster to operate. Given that the configuration of the PRISM large language model involves two additional interceptor models $p$ and $q$, it is essential that $p$ and $q$ are significantly smaller in size than the underlying language model $f$. 

The interceptor models $p$ and $q$ can be trained to learn and enforce AUPs from the underlying large language model $f$ using a technique called knowledge distillation \citep{hinton2015distilling}. Knowledge distillation is a process in which a smaller model (the student) is trained to mimic the behavior of a larger, more complex model (the teacher) by learning from the teacher's outputs \citep{hinton2015distilling}.

In this case, the large language model $f$ serves as the teacher, and the interceptor models $p$ and $q$ act as the students. By exposing $f$ to a diverse range of prompts and analyzing its outputs, the developers can identify patterns and behaviors that align with or violate the desired AUPs. This information can then be used to create a labeled dataset, along with developer community feedback. The interceptor models $p$ and $q$ are trained on this labeled dataset, learning to predict whether a given input or output sequence is ``unsafe'', or likely to violate the AUPs. By minimizing the difference between their predictions and the labels derived from the large language model's behavior, the interceptor models effectively distill the knowledge about AUPs from the teacher model into a more compact and computationally efficient form.

\paragraph{Utility Gains for End-Users}

When comparing a more efficient model to a less efficient model holding utility constant, end-users directly benefit from cost savings for compute and faster model performance. These two factors make minimizing the marginal compute of safety mechanisms a worthwhile goal for model developers. 

\paragraph{Utility Gains for Society-at-Large}

Energy dedicated to running and processing foundation models is increasing every day \citep{wef2024manage}. Larger models consume more energy per inference than smaller, more efficient models. Therefore, optimizing a model to be as efficient offers huge potential for energy savings, especially considering that the foundation model will be used by thousands, if not millions, of developers throughout the world.

\section{Discussion}
\subsection{Summary of Findings}

\paragraph{Open Language Models are Improving at the Same or Greater Rate than Closed Language Models}

Open-source foundation models are rapidly advancing, with their capabilities improving at a rate equal to or faster than their closed-source counterparts. This accelerated progress, driven by the collective efforts of a global research community, positions open-source models as a dominant force in the AI landscape. As these models become more powerful and widely adopted, their potential for both beneficial applications \emph{and} harmful misuse grows.

\paragraph{Open Foundation Model Companies Provide Less Restrictive AUPs than Closed Model Companies.}

Our analysis reveals that open-source foundation model companies often provide less restrictive acceptable use policies (AUPs) compared to closed-source model developers. This discrepancy may stem from the inherent challenges in enforcing usage restrictions once the models are released or from the open-source philosophy that values accessibility and innovation over strict limitations. Consequently, open-source models may be more vulnerable to misuse by malicious actors.

\paragraph{The PRISM Framework Offers a Solution to Vulnerabilities in Open Foundation Models.}

By introducing modular interceptor functions that independently moderate prompts and outputs, PRISM provides a robust defense against common attacks such as prompt injection and malicious fine-tuning. This adaptable approach focuses on identifying AUP violations rather than directly modifying the language model, ensuring a more resilient safety mechanism.

\paragraph{The PRISM Framework Offers Utility Gains for End-Users and Society.}

For end-users, PRISM enables the use of sensitive data while maintaining privacy and provides greater control over AUP enforcement. Minimizing the cost of compute for the interceptor functions $p$ and $q$ makes these models more desirable to end-users as well.  On a societal level, the framework helps address public concerns about the potential misuse of open-source foundation models by incorporating diverse perspectives and ensuring responsible development practices.

\subsection{Limitations}

While the PRISM framework offers a promising approach to open-source foundation model safety, its real-world effectiveness remains to be empirically tested. The proposed interceptor functions have not yet been implemented and evaluated in practical scenarios, leaving questions about their performance and reliability unanswered. Additionally, the study does not provide a comprehensive analysis of the potential computational overhead introduced by these interceptor functions. 

Furthermore, the long-term adaptability of PRISM to the rapidly evolving AI safety landscape requires further investigation. As new threats and challenges emerge, the framework may need to be continually updated and refined to remain effective. This adaptability is particularly crucial in light of the increasing capabilities of open-source foundation models and the potential for malicious actors to exploit vulnerabilities in novel ways.

Lastly, the study acknowledges that some exemptions from safety measures may be necessary for research purposes. In order to foster innovation and progress in the field of AI, researchers may require access to models without certain restrictions to investigate novel approaches and test hypotheses.

\subsection{Next Steps}

In a future version of the paper, we will build a large language model using the PRISM framework and empirically test the extent to which this model is (i) more resistant to prompt injection and fine-tuning attacks, (ii) able to remotely enforce AUPs in various contexts, and (iii) perceived as more useful to end-users.

\clearpage
\bibliographystyle{apacite}
\bibliography{sample}

\end{document}